\documentclass{PoS}
\usepackage{amsfonts}
\usepackage{amsmath}
\usepackage{subfigure} 

\graphicspath{{./figs/}}

\title{
  Update on $B_K$ and $\varepsilon_K$ with staggered quarks 
}

\ShortTitle{
  Recent update on $B_K$
}

\author{Taegil Bae, Yong-Chull Jang, Hwancheol Jeong, Jangho Kim, 
  Jongjeong Kim, Kwangwoo Kim, Seonghee Kim, 
  \speaker{Weonjong Lee}, Jaehoon Leem, 
  Jeonghwan Pak, Sungwoo Park \\
  Lattice Gauge Theory Research Center, CTP, and FPRD, \\
  Department of Physics and Astronomy,
  Seoul National University, Seoul, 151-747, South Korea \\
  E-mail: \email{wlee@snu.ac.kr}}

\author{Chulwoo Jung, Hyung-Jin Kim \\
  Physics Department, Brookhaven National Laboratory,
  Upton, NY11973, USA \\
  E-mail: \email{chulwoo@bnl.gov}}

\author{Stephen R. Sharpe\\
  Physics Department, University of Washington, 
  Seattle, WA 98195-1560, USA \\
  E-mail: \email{sharpe@phys.washington.edu}}

\author{Boram Yoon \\
  Los Alamos National Laboratory, \\
  Theoretical Division T-2, MS B283, \\
  Los Alamos, NM 87545, USA \\
  E-mail: \email{googlus@gmail.com}}

\author{SWME Collaboration}

\abstract{ We update our results for $B_K$ obtained using
  HYP-smeared staggered valence quarks on the MILC asqtad lattices. 
  In the last year, we have
  added 5 new measurments on the fine ($a\approx 0.09\;$fm)
  ensembles, and 2 new measurements on the superfine ($a\approx 0.06\;$fm)
  ensembles. These allow a simultaneous extrapolation in
  $a^2$ and sea quark masses, reducing the corresponding
  systematic error  significantly. Our updated result is 
  $\hat{B}_K = 0.738 \pm 0.005 (\text{stat}) \pm 0.034 (\text{sys})$.
}

\FullConference{ 31st International Symposium on Lattice Field Theory
  - LATTICE 2013\\ July 29 - August 3, 2013\\ Mainz, Germany }

\begin{document}

\section{Introduction} 
The standard model prediction for $\varepsilon_K$ is
proportional to the kaon mixing matrix element parametrized
by $B_K$. Recent progress in the calculation of $B_K$ and
other quantities using lattice QCD~\cite{Colangelo:2010et} 
allows a high-precision test of the standard model.
Although $B_K$ is a subdominant source of error in present estimates
of $\varepsilon_K$, this may well change in the future,
so further reduction in the errors is worthwhile.

Here we update the determination of $B_K$ 
using improved staggered quarks. At Lattice 2012, we found the
surprising result that the slope of $B_K$ versus light sea-quark mass
depended non-monotonically on the lattice spacing 
(see Fig.~3 of Ref.~\cite{Bae:2012um}). 
To investigate further, we have
added 7 new lattice ensembles with different
values of the sea-quark masses (see Table~\ref{tab:milc-lat}). 
This has resolved last year's problem, as described below.

Table~\ref{tab:milc-lat} lists all the MILC asqtad ensembles on
which we have calculated (or are calculating) $B_K$.
In our earlier calculations, we used only a subset of these
ensembles. Initially, we took the continuum limit
using ensembles F1, S1 and U1 (i.e. holding the ratio of 
light to strange sea-quark masses fixed), while estimating the
sea-quark mass dependence from the coarse 
lattice ensembles C1-C5~\cite{Bae:2011ff}.
By Lattice 2012, we had added ensembles F2, F3, S2 and S3
(and increased statistics on several ensembles)~\cite{Bae:2012um}.
Since then we have added measurements on
ensembles F4, F5, F6, F7, F9, S4 and S5 
(with F8 and S6 in the pipeline).
The net effect is that we can study the sea-quark mass
dependence in much greater detail, and in particular do a
combined continuum, light sea-quark mass 
and strange sea-quark mass extrapolation. 

\begin{table}[h!]
\begin{center}
\begin{tabular}{ c | c | c | c | c | c }
\hline $a$ (fm) & $am_l/am_s$ & geometry & ID & ens $\times$ meas
& status \\
\hline
0.12 & 0.03/0.05  & $20^3 \times 64$ & C1 & $564 \times 9$ & old \\
0.12 & 0.02/0.05  & $20^3 \times 64$ & C2 & $486 \times 9$ & old \\
0.12 & 0.01/0.05  & $20^3 \times 64$ & C3 & $671 \times 9$ & old \\
0.12 & 0.01/0.05  & $28^3 \times 64$ & C3-2 & $275 \times 8$ & old \\
0.12 & 0.007/0.05 & $20^3 \times 64$ & C4 & $651 \times 10$ & old \\
0.12 & 0.005/0.05 & $24^3 \times 64$ & C5 & $509 \times 9$ &  old \\
\hline
0.09 & 0.0062/0.0186 & $28^3 \times 96$ & F6 & $950 \times 9$ & \texttt{new} \\
0.09 & 0.0124/0.031 & $28^3 \times 96$ & F4 & $1995 \times 9$ & \texttt{new} \\
0.09 & 0.0093/0.031 & $28^3 \times 96$ & F3 & $949 \times 9$  & old \\
0.09 & 0.0062/0.031 & $28^3 \times 96$ & F1 & $995 \times 9$  & old \\
0.09 & 0.00465/0.031 & $32^3 \times 96$ & F5 & $651 \times 9$ & \texttt{new} \\
0.09 & 0.0031/0.031 & $40^3 \times 96$ & F2 & $959 \times 9$  & old \\
0.09 & 0.0031/0.0186 & $40^3 \times 96$ & F7 & $701 \times 9$ & \texttt{new} \\
0.09 & 0.0031/0.0031 & $40^3 \times 96$ & F8 & $576 \times 9$ & NA \\
0.09 & 0.00155/0.031 & $64^3 \times 96$ & F9 & $790 \times 9$ & \texttt{new} \\
\hline
0.06 & 0.0072/0.018  & $48^3 \times 144$ & S3 & $593 \times 9$ & old \\
0.06 & 0.0054/0.018  & $48^3 \times 144$ & S4 & $582 \times 9$ & \texttt{new} 
\\
0.06 & 0.0036/0.018  & $48^3 \times 144$ & S1 & $749 \times 9$ & old \\
0.06 & 0.0025/0.018  & $56^3 \times 144$ & S2 & $799 \times 9$ & old \\
0.06 & 0.0018/0.018  & $64^3 \times 144$ & S5 & $821 \times 6$ & \texttt{new} 
\\
0.06 & 0.0036/0.0108 & $64^3 \times 144$ & S6 & $600 \times 0.05$ & NA \\
\hline
0.045 & 0.0028/0.014 & $64^3 \times 192$ & U1 & $747 \times 1$ & old \\
\hline
\end{tabular}
\end{center}
\caption{MILC asqtad ensembles used to calculate $B_K$.  $a m_\ell$
  and $a m_s$ are the masses, in lattice units, of the light and
  strange sea quarks, respectively. ``ens'' indicates the number of
  configurations on which ``meas'' measurements are made.  Note that
  the numbering of the ID tags on the fine and superfine lattices do
  not follow the ordering of $a m_\ell$.  ``NA'' means that
  analysis results are not yet available.}
\label{tab:milc-lat}
\end{table}

\section{Valence quark mass extrapolations}
We used a mixed action, with asqtad sea quarks and
HYP-smeared~\cite{Hasenfratz:2001hp} valence quarks.
We denote the masses of the valence $d$ and $s$ quarks by 
$m_x$ and $m_y$, respectively, while the
light and strange sea-quark masses are $m_\ell$ and $m_s$.
On each ensemble, we use 10 valence masses:
$am_x, am_y = am_s^{\rm nom} \times (n/10)$ with $n = 1,2,3,\ldots,10$,
where $am_s^{\rm nom}=0.05$, $0.030$, $0.018$ and $0.014$ on the
coarse, fine, superfine and ultrafine ensembles, respectively.
We extrapolate to the physical value of $m_d$ using the
lightest four values of $m_x$,
and to the physical $m_s$ using the heaviest three $m_y$.
We are then in the regime ($m_x\ll m_y \sim m_s$) where SU(2) 
[staggered] chiral perturbation theory ([S]ChPT) is applicable.
%

\begin{figure}[t!]
\subfigure[F9]{\includegraphics[width=0.49\textwidth]
  {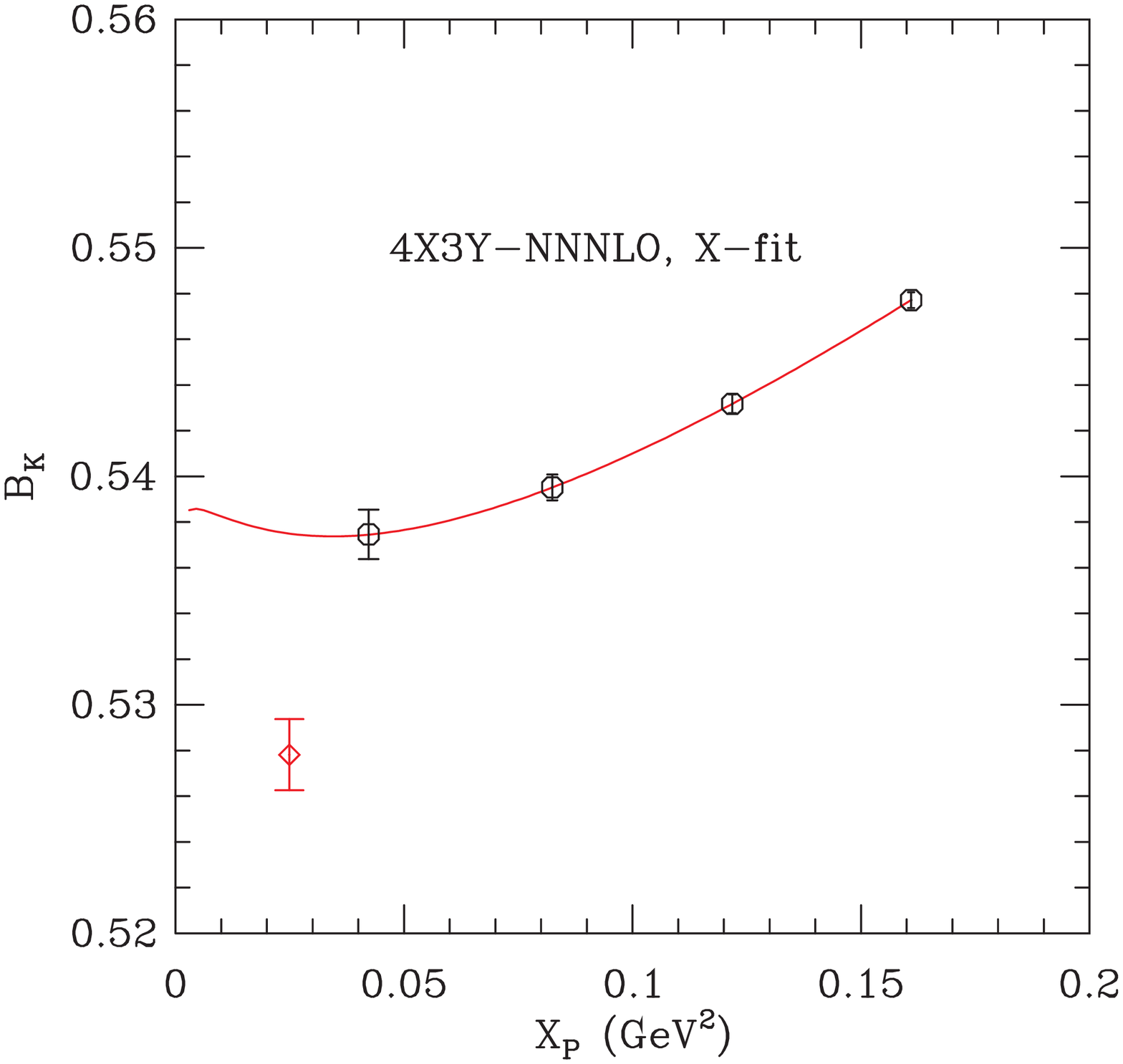}}
\subfigure[S5]{\includegraphics[width=0.49\textwidth]
  {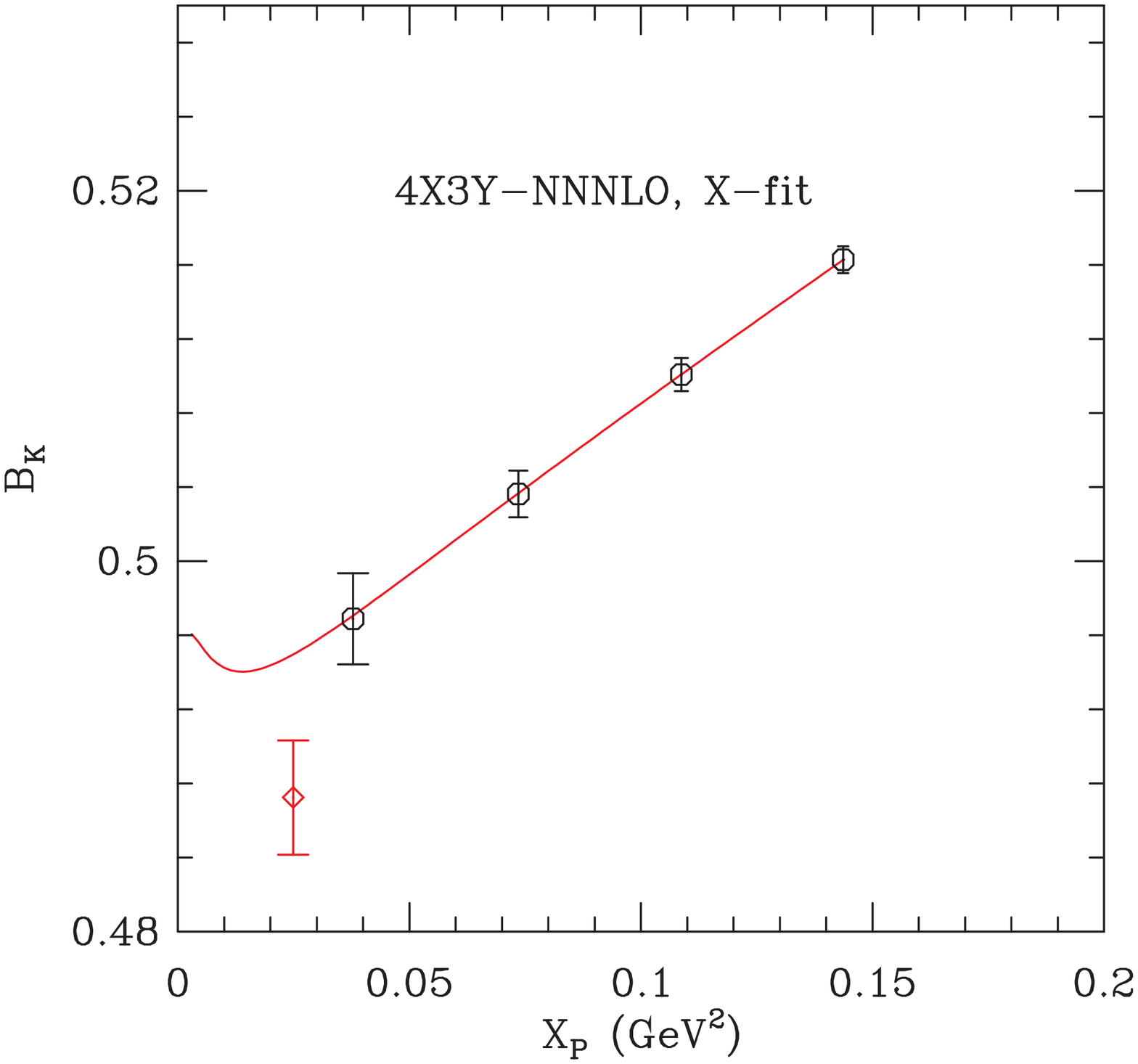}}
\caption{X-fits to $B_K(1/a)$ for the F9 and S5 ensembles. 
Here $X_P$ is the mass of the $x\bar x$ valence taste-$\xi_5$ pion.
The red diamond is explained in the text.}
\label{fig:X-fit:F9+S5}
\end{figure}

We call the extrapolation in $m_x$ the ``X-fit''.
We fit to the next-to-leading order (NLO) SChPT finite-volume
form worked out in 
Refs.~\cite{VandeWater:2005uq,Bae:2010ki},
augmented by NNLO and higher order terms,
including Bayesian constraints,
as described in Refs.~\cite{Bae:2010ki,Bae:2011ff}.
Examples of these fits for 
two of the new ensembles 
are shown in Fig.~\ref{fig:X-fit:F9+S5}.
These are the ensembles with the lightest light sea quarks
at the ``fine'' ($a\sim 0.09\;$fm) and
``superfine'' ($a\sim 0.06\;$fm) lattice spacings.
Indeed, on ensemble F9 our sea quarks have $m_\ell=m_s/20$,
which is lighter than our lightest valence quark,
and corresponds to a sea-quark pion of mass $\sim 180\;$MeV.
In the figures, the red diamond is the value
obtained after extrapolating to $m_x=m_d$,
setting the pion masses appearing in the NLO chiral logarithms
to their physical values (with taste-breaking removed)
and setting the volume to infinity.
Systematic errors in the X-fits are estimated by varying the
Bayesian priors and by using fits with and without NNNLO terms.

The extrapolation of $m_y$ to the physical $m_s$ 
(the ``Y-fit'') is done using
linear and quadratic fits. The quadratic terms are very small,
as in our earlier work~\cite{Bae:2011ff,Bae:2010ki}.
We use the linear fits for the central value and the quadratic
fits to estimate a systematic error.
\section{Continuum Extrapolation}
\label{sec:am_l}
At this stage, we have one-loop matched results
for $B_K(1/a)$ on each ensemble. We first run these to a common
scale, which we take to be $2\;$GeV.
The remaining errors are those due to discretization 
(primarily taste-conserving), the need to extrapolate in 
the sea-quark masses $m_\ell$ and $m_s$, and truncation errors
in the matching factors.
Note that the sea-quark mass dependence is analytic at NLO, because
we have accounted for the chiral logarithms in the valence-quark
extrapolations.

With our much enlarged data-set, it is now 
possible to perform a simultaneous fit to $a^2$, $m_\ell$ and
$m_s$, which is a significant improvement compared to our previous work.
We have tried a number of fit functions, but discuss here only the
simplest and most complicated forms, which we label B1 and B4, respectively.
The B1 fit function is
\begin{align}
  f_\text{B1} &= c_1 + c_2 (a\Lambda_Q)^2 
  + c_3 \frac{L_P}{\Lambda_X^2} + c_4 \frac{S_P}{\Lambda_X^2} \,.
\end{align}
with $L_P$ ($S_P$) the squared pion masses of the 
taste-$\xi_5$ $\ell\bar{\ell}$ ($s\bar{s}$) pions.
The scales are chosen to be
$\Lambda_Q = 0.3\;$GeV and $\Lambda_X = 1.0\;$GeV,
with Bayesian constraints $c_i = 0 \pm 2$ for $i=2,3,4$.
This forces the parameters to have magnitudes similar to those expected
from dimensional analysis.
The linear dependence on $L_P$ is the prediction of NLO SChPT,
while that on $S_P$ is just the simplest choice for a smooth function.

\begin{figure}[t!]
\centering
\includegraphics[width=0.7\textwidth]{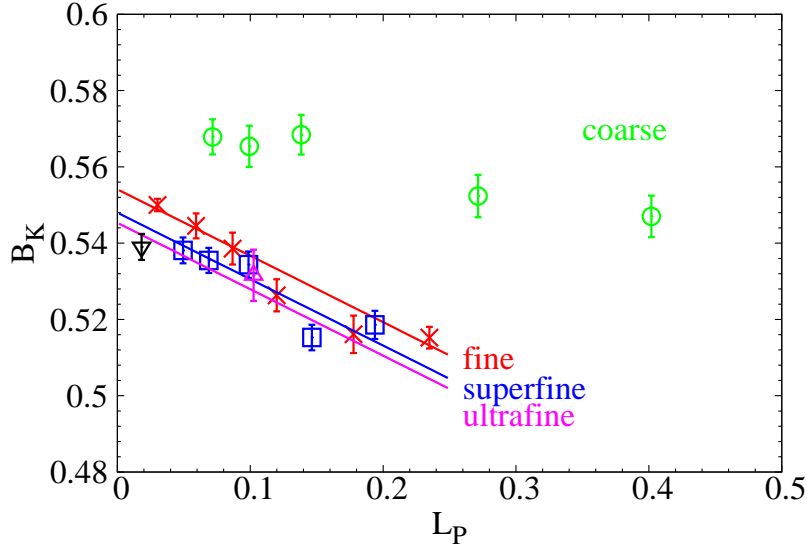}
\caption{$B_K(\mu=2\,\text{GeV})$ vs. $L_P \, (\text{GeV}^2)$ with
a B1 fit. The black diamond is result at $a=0$ and with physical
sea-quark masses. The lines show the fit function with
$a^2$ and $S_P$ fixed to the average value for the corresponding ensembles
(fine, superfine or ultrafine), except that for the fine ensembles
$S_P$ is the average of the values on ensembles F1, F2 and F4.
}
\label{fig:bk-su2-cont-ext:B1}
\end{figure}
We show the B1 fit in Fig.~\ref{fig:bk-su2-cont-ext:B1}.
Although results from the coarse ensembles C1-C5 are displayed, 
they are not included in the fit.
Doing so leads to very low confidence levels for all fit
forms we have tried.
Thus we include in the fit only the 8 fine, 5 superfine and 1 ultrafine
ensembles, and find a reasonable fit with $\chi^2/\text{dof} = 1.46$.
We note that the fine and superfine
points should not lie precisely on the corresponding lines shown in the plots,
because their values for $a^2$ and $S_P$ vary slightly 
(by up to $6\%$ and $3\%$, respectively).
This discrepancy is much larger for ensembles F6 and F7,
which have significantly different values of $am_s$, and
so we do not display the results from these two ensembles
(although they are included in the fit).
These ensembles give us a strong ``lever-arm'' for determining the
$S_P$ dependence.
We stress that we do not build in SU(3) symmetry---$c_3$ and $c_4$
are independent parameters, and indeed turn out to differ
significantly.

\begin{figure}[t!]
\centering
\includegraphics[width=0.7\textwidth]{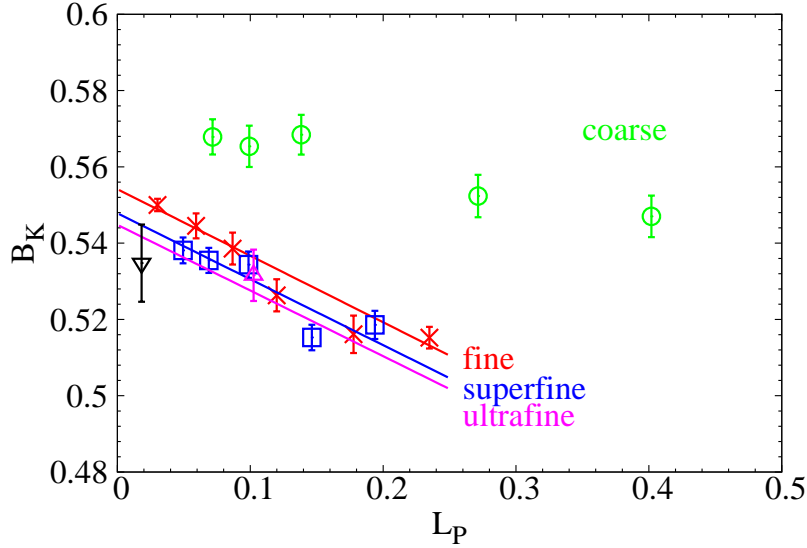}

\caption{$B_K(\mu=2\,\text{GeV})$ vs. $L_P \, (\text{GeV}^2)$ with
fit function B4.
}
\label{fig:bk-su2-cont-ext:B4}
\end{figure}

The B4 fit uses the form
\begin{align}
  f_\text{B4} &= f_\text{B1} 
+ c_5 (a\Lambda_Q)^2 \frac{L_P}{\Lambda_\chi^2}
+ c_6 (a\Lambda_Q)^2 \frac{S_P}{\Lambda_\chi^2}
+ c_7 [\alpha_s(\frac1a)]^2 
+ c_8 (a\Lambda_Q)^2 \alpha_s(\frac1a)
+ c_9 (a\Lambda_Q)^4
\end{align}
The most significant new term is that proportional to $\alpha_s^2$,
since this varies the most slowly with $a$.
This term is present because we use one-loop matching.
All the new terms are constrained along the lines described above.
The result of the B4 fit is shown in
Fig.~\ref{fig:bk-su2-cont-ext:B4}.
The quality of fit barely changes from the B1 fit, 
with $\chi^2/\text{dof}$ still $1.46$.
The main change in the B4 fit is an increase in
the statistical error (as we expect with more parameters), 
along with a shift in the central value
which is not statistically significant.
We see that the data neither ``wants'' nor excludes the extra terms
in the fit function.
Nevertheless, since the extra terms are theoretically well motivated,
we use the difference between the results of the B4 and B1 fits as
our estimate of the systematic error in the continuum-chiral extrapolation,
while using B1 for the central value.

As mentioned in the introduction,
our results last year showed a non-monotonicity in the 
dependence of the slopes versus $L_P$ as we approached the
continuum limit. Comparing to
Fig.~3 of Ref.~\cite{Bae:2012um}, we find that two factors
contribute to the resolution of this problem.
First, adding more values of $L_P$ allows the
slopes to be better determined, and we then find that they are
consistent with monotonic dependence on $a$.
Second, we allow for independent $L_P$ and $S_P$ dependence,
and account for the variation in $a^2$ and $S_P$ between ensembles.

\section{Final Result and Outlook}

After extrapolation we find
%
%
\begin{align}
\hat{B}_K &= 0.738 \pm 0.005 (\text{stat}) \pm 0.034 (\text{sys}) 
\end{align}
The sources of error and their contributions are
collected in Table~\ref{tab:err-budget}.
Our methods for estimating the main systematic
errors have been described above,\footnote{%
We estimate minor errors following the
methods described in Ref.~\cite{Bae:2010ki}.}
with the exception of the matching factor error.
The latter arises from truncating the perturbative matching factor at
one-loop order.
We estimate the resulting error as $\Delta B_K/B_K = \alpha_s^2$ with $\alpha_s$
evaluated at scale $1/a$ on the finest (U1) lattice.
We note that the difference between B4 and B1 fits includes,
in part, an estimate of this truncation error. 
Thus, when we combine all errors in quadrature, there is some double
counting. This is numerically a small effect, however, and
we ignore it.

\begin{table}[tbhp]
\begin{center}
\begin{tabular}{ l | l l }
  \hline
  \hline
  cause & error (\%) & memo \\
  \hline
  statistics         & 0.63  & see text \\
  matching factor    & 4.4   & $\Delta B_K^{(2)}$ (U1) \\
  $ \left\{ \begin{array}{l} 
    \text{discretization} \\ a m_\ell \text{ extrap} \\ a m_s \text{ extrap}
  \end{array} \right\} $
                     & 1.1   & diff.~of B1 and B4 fits \\
  X-fits             & 0.33  & varying Bayesian priors (S1) \\
  Y-fits             & 0.53  & diff.~of linear and quad. (F1) \\
  finite volume      & 0.5   & diff.~of $V=\infty$ and FV fit~\cite{Kim:2011qg}\\
  $r_1$              & 0.27  & $r_1$ error propagation (F1) \\
  $f_\pi$            & 0.4   & $132\;$MeV vs. $124.4\;$MeV \\
  \hline
  \hline
\end{tabular}
\end{center}
\caption{Error budget for $B_K$ using SU(2) SChPT fitting. 
  \label{tab:err-budget}}
\end{table}

Our final result is completely consistent with that we
found previously using many fewer ensembles (C1-5, F1, S1 and U1),
namely $\hat B_K=0.727(4)(38)$~\cite{Bae:2011ff}.
The extra ensembles have led to a substantial reduction in
the errors from continuum and sea-quark mass extrapolations:
this error was previously 2.7\% and is now 1.1\%.
This improvement only leads to a small reduction in the total
systematic error, however, due to the dominant (and unchanged)
matching error. 

As in Ref.~\cite{Jang:2012ft}, we can convert the above results
into predictions for $\varepsilon_K$.
Preliminary results\footnote{The final results will be reported
in Ref.~\cite{wlee2013jyc}.} are
\begin{align}
|\varepsilon_K| &=  1.51(18) \times 10^{-3}\qquad \text{for exclusive $V_{cb}$}
\\
|\varepsilon_K| &=  1.91(21) \times 10^{-3} \qquad \text{for inclusive $V_{cb}$}
\,.
\end{align}
The former value lies $4 \sigma$ away from the experimental value
$|\varepsilon_K| = 2.228(11) \times 10^{-3}$.
Further improvement clearly requires reducing the matching factor
error.
To do so we are
calculating the matching factors using non-perturbative
renormalization (NPR) in the RI-MOM and RI-SMOM schemes.
Preliminary results (for bilinears) are reported in Ref.~\cite{Kim:2013bta}.
See also Ref.~\cite{Lytle:2013qoa}.
We expect that NPR will reduce the error in matching down to the $\sim 2\%$ 
level.
We are also pursuing a two-loop perturbative matching calculation.

\section{Acknowledgments}
We are grateful to Claude Bernard and the MILC collaboration
for private communications.
C.~Jung is supported by the US DOE under contract DE-AC02-98CH10886.
The research of W.~Lee is supported by the Creative Research
Initiatives Program (2013-003454) of the NRF grant funded by the
Korean government (MSIP).
W.~Lee would like to acknowledge the support from KISTI supercomputing
center through the strategic support program for the supercomputing
application research [No.~KSC-2012-G3-08].
The work of S.~Sharpe is supported in part by the US DOE grant
no.~DE-FG02-96ER40956.
Computations were carried out in part on QCDOC computing facilities of
the USQCD Collaboration at Brookhaven National Lab, on GPU computing
facilities at Jefferson Lab, on the DAVID GPU clusters at Seoul
National University, and on the KISTI supercomputers. The USQCD
Collaboration is funded by the Office of Science of the
U.S. DOE.
%

\bibliographystyle{JHEP}
\bibliography{ref}

\end{document}